\documentclass[conference]{IEEEtran}
\IEEEoverridecommandlockouts

\usepackage[numbers,sort&compress]{natbib}
\usepackage{amsmath,amssymb,amsfonts}
\usepackage{algorithmic}
\usepackage{graphicx}
\usepackage{textcomp}
\usepackage{xcolor}
\def\BibTeX{{\rm B\kern-.05em{\sc i\kern-.025em b}\kern-.08em
    T\kern-.1667em\lower.7ex\hbox{E}\kern-.125emX}}

\usepackage{booktabs}
\usepackage{multirow}

\usepackage{tikz}
\usepackage{comment}
\usepackage[utf8]{inputenc}

\usetikzlibrary{arrows.meta, shapes}
\usetikzlibrary{positioning}
\usepackage{hyperref}
\hypersetup{
    colorlinks   = true,
    citecolor    = blue,
    urlcolor    =  blue
}

\newcommand{\hf}[2]{\raisebox{-2.2pt}{\includegraphics[scale=0.09]{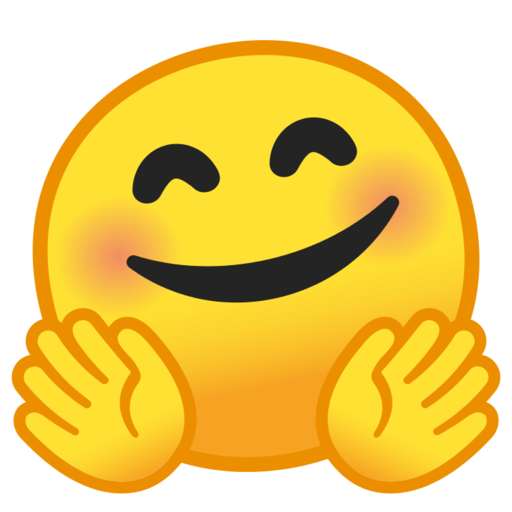}}~\href{#1}{\texttt{#2}}}

\newcommand{\gh}[2]{\raisebox{-2.2pt}{\includegraphics[scale=0.02]{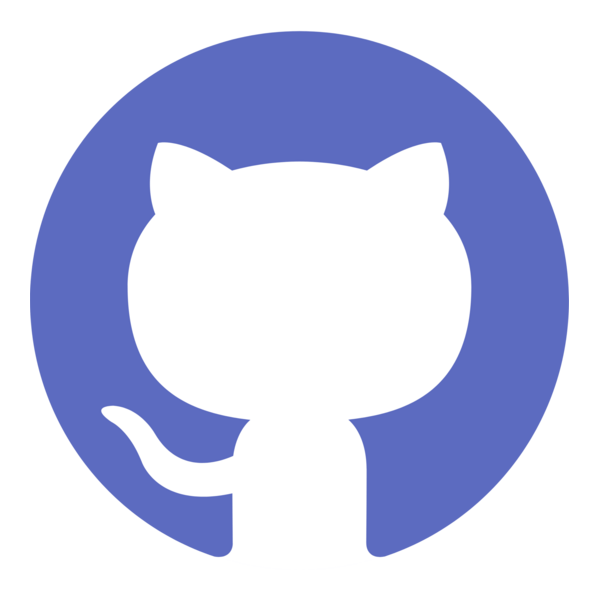}}~\href{#1}{\texttt{#2}}}

\begin{document}

\title{Forecasting Weather-Driven Price Dynamics Across Sri Lankan Tea Market Catalogues}

\author{
\IEEEauthorblockN{%
Hesandi Mallawarachchi,
Senilka Madurapperumage,
Nadil Kulathunge,
Thilokya Angeesa,
Nethsith Gunaweera,\\
Sandeepa Weerasekara,
Patalee Narasinghe,
Nisansa de Silva,
Sandareka Wickramanayake}
\IEEEauthorblockA{Dept.\ of Computer Science \& Engineering, University of Moratuwa, Sri Lanka.\\
\texttt{\{hesandim.23, senilkam.23, nadilk.23, thilokyaa.23, nethsithg.23,}\\
\texttt{sandeepa.25, patalee.21, NisansaDds, sandarekaw\}@cse.mrt.ac.lk}
}
}

\maketitle

\begin{abstract}

We construct and publicly release a novel structured dataset of 12,233 records parsed from 105 weekly broker PDF reports spanning late 2023 to 2026, a machine-readable archive of Colombo Tea Auction (CTA) prices combined with region-specific lagged weather variables from the Open-Meteo historical archive. Using this dataset, we investigate the relationship between local weather conditions and price behavior across four Sri Lankan tea catalogues: High Grown, Low Grown, Off-Grade, and Dust. We apply Granger causality analysis alongside four tree-based machine learning models: \texttt{Random Forest}, \texttt{XGBoost}, \texttt{LightGBM}, and \texttt{Gradient Boosting}. Our results show that market dynamics are primary drivers, while weather provides limited catalogue-specific predictive information. Extended Granger analysis across cumulative lag orders of 1--8 weeks identified one FDR-significant association: temperature history through four weeks Granger-predicted Dust prices, while other nominal weather associations were treated as exploratory. Catalogue-specific modelling outperformed unified approaches, with \texttt{LightGBM} emerging as the superior model for three out of four catalogues, though Naive Persistence was the strongest overall method for Low Grown.
Overall, this study highlights the importance of considering both localized weather patterns and catalogue-level differences when forecasting tea prices, offering a more precise and practical framework for the tea industry.
\end{abstract}


\section{Introduction}

As of 2024, Sri Lanka was the world's fifth largest tea producer~\cite{sltb2024annual} and remains a leading exporter of orthodox (Ceylon) tea, with tea contributing roughly 10\% of the country's agricultural export earnings and supporting more than two million livelihoods among smallholders, estate workers, and downstream trades~\cite{board2024sri}. The Colombo Tea Auction (CTA), held weekly by the Tea Board of Sri Lanka, is the primary price-discovery mechanism for Ceylon tea and sets benchmarks that flow into forward contracts and retail shelves in more than 90 countries. Price signals from the CTA therefore have direct welfare consequences for a large, weather-exposed workforce.

\subsection{Tea Categorization in Sri Lanka}

Sri Lankan tea is primarily categorized using two complementary systems\footnote{Information based on weekly market reports by \href{{https://www.forbestea.com/}}{Forbes \& Walker Tea Brokers (Pvt) Ltd}, the leading tea broker in Sri Lanka} that together provide both geographical and commercial context for auction pricing and market analysis.

\subsubsection{Categorization by Elevation (Geographical/Regional)}

This is the most structurally important grouping for price modeling and supply analysis, as elevation strongly influences tea quality, flavor profile, and growing conditions.

\begin{itemize}
\item \textbf{High Grown Areas} - Teas cultivated at higher altitudes, typically above $\sim$1,200 m (e.g., Nuwara Eliya, Western High, Uda Pussellawa, Uva regions).
\item \textbf{Medium Grown Areas} - Teas from mid-elevation estates, roughly 600 - 1,200 m.
\item \textbf{Low Grown Areas} - Teas grown below $\sim$600 m, mainly in the southern and southwestern regions (Galle, Matara, Ratnapura).
\end{itemize}

\subsubsection{Categorization by Catalogue Type}

This mainly reflects tea-leaf quality and how teas are physically sorted, processed and presented at the Colombo Tea Auction:

\begin{itemize}
\item \textbf{High Grown Catalogue} - Primarily consists of High and Medium Grown teas sold in bulk under individual estate names.
\item \textbf{Low Grown Catalogue} - The high-volume Low Grown production, further subdivided by leaf size and style. These teas are highly prized for their visual appeal and brisk taste.
\item \textbf{Off-Grade} - Teas that fall outside standard leaf grades during processing, often containing higher proportions of fiber, stalk, or broken leaves.
\item \textbf{Dust} - The smallest particle size fraction, consisting of the finest tea dust.
\end{itemize}

Importantly, both Off-Grade and Dust are mixed categories that include teas from all three growing regions: High, Medium, and Low Grown. These four catalogues show substantial differences in average auction prices, yet most prior 
computational work on Sri Lankan tea prices pools all grades into a single target or uses annual aggregates~\cite{pasandul2026tea}, making it impossible 
to recover the catalogue-level drivers that brokers actually use for decision-making.

Another practical challenge is data. CTA price information is distributed as weekly PDF reports issued by the eight licensed brokers including semi-structured tables buried in prose commentary and there is no public machine readable archive. By contrast, weather data is abundant but must be aligned to the correct growing region, auction week and linked through the biological lag between leaf formation and sale. No prior study has combined a machine parse-able CTA corpus with region specific weather and formally tested whether weather Granger causes~\cite{granger1969investigating} catalogue-level prices in the sense of~\cite{aslan2024practical}.

This paper makes two contributions to address that gap:
\begin{itemize}
    \item A novel structured dataset of price observations parsed from 105 weekly Forbes and Walker broker reports (November 2023 - March 2026) enriched with region-specific daily weather from the Open-Meteo historical archive and aligned at the sale week level with lagged weather features.
        
    \item Practical guidance on catalogue specific price drivers. This allows brokers to provide more accurate market advice and buyers to make better sourcing and bidding decisions.
\end{itemize}

\section{Related Work}

\textbf{Tea Price Forecasting.}
Most earlier studies on Sri Lankan tea prices have taken a
macroeconomic approach \cite{de2022export}, examining how exchange
rates, oil prices and supply from other countries affect
aggregate export prices \cite{rupasinghe2023investigation}. These analyses
operate at monthly or annual frequency and pool across all
grades, which makes them well suited to trade-balance questions.
However, those analyses are unable to resolve the weekly, catalogue-level dynamics on which brokers and buyers actually transact. More recently, researchers have started using machine learning methods such as \texttt{ARIMA}, \texttt{SVR}, and \texttt{LSTM} on Indian tea auction data \cite{chaundry2017time}, and a few attempts have been made to model Sri Lankan prices as well \cite{pasandul2026tea}. However, these studies usually treat all teas as one large group. Most studies do not explicitly model heterogeneity across grades and auction catalogues, and few studies incorporate region specific weather variables into tea price forecasting \cite{chaundry2017time}. This leaves a clear gap because it is not yet clear how weather and tea-quality interact across the four distinct catalogues in the Colombo Tea Auction.

\textbf{Time-Series Modeling of Agricultural Commodities.}
Temporal dependency is a key challenge in forecasting agricultural commodity prices due to the volatile nature of price series and sensitivity to external shocks. As a result, recent studies incorporate temporal structures using machine learning and deep learning approaches to improve forecasting performance \cite{rl2021forecasting}. However, limited work has examined weekly tea auction prices in Sri Lanka using time series techniques, particularly in relation to lagged weather effects across different market catalogues.

\textbf{Weather and Price Linkages.}
Agronomists have long shown that rainfall, temperature, and sunshine strongly influence tea yield \cite{wijeratne1996vulnerability} and leaf quality in Sri Lanka \cite{ahmed2018global}. Although prior agronomic studies in Sri Lanka have examined how weather affects tea yield and quality, those findings have not been directly tested against weekly, catalogue-level auction price formation in the Colombo Tea Auction. There remains a clear gap between agronomic evidence and catalogue-level price behavior at auction.

\section{Data Sources}
The main dataset is constructed from weekly market reports published by 
Forbes \& Walker Tea Brokers (Pvt) Ltd, one of eight licensed brokers 
operating at the Colombo Tea Auction and a member of the Colombo Tea 
Traders' Association (CTTA). These reports are publicly 
available \cite{board2024sri}. A total of 105 weekly reports from November 
2023 to March 2026 were collected for initial analysis\footnote{\url{https://www.forbestea.com/statistics-tea-market-reports}}.

Meteorological data was obtained from the \textbf{Open-Meteo historical 
archive API}~\cite{zippenfenig_2024_14582479}, which is free and publicly accessible and 
provides daily observations without authentication. Weather variables, 
including total precipitation (mm), mean temperature ($^\circ$C) and 
sunshine duration (seconds), were obtained for tea-growing coordinates 
representing High Grown, Medium, and Low Grown regions. \footnote{Region centroids used 
for Open-Meteo API calls: Western High Grown/Maskeliya 
(6.9271°N, 80.5350°E), Nuwara Eliya (6.9497°N, 80.7891°E), 
Uva/Uda Pussellawa (6.8700°N, 81.0600°E), Low Grown/Matara--Galle 
(6.2500°N, 80.3000°E). Off-Grade and Dust catalogues use Low Grown 
as primary and Western High Grown as secondary region, reflecting 
their mixed-elevation sourcing.}
Key features 
extracted from the sale reports include catalogue average price in LKR and 
USD, total volume sold (kg), text-parsed weather condition per region, and 
crop intake direction.

\section{Methodology}
The overall methodology follows a five-stage end-to-end pipeline as 
displayed in Figure~\ref{fig:methodology_pipeline}:

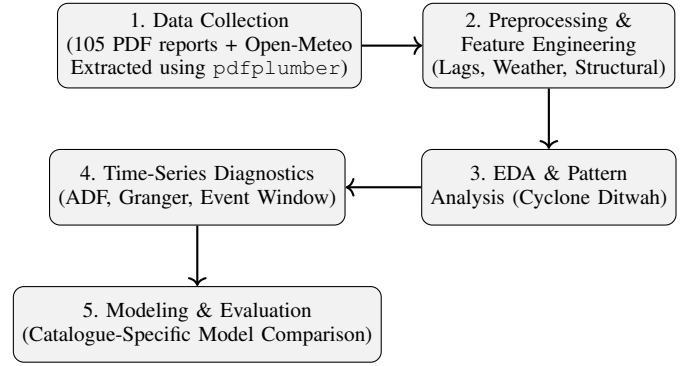
\begin{figure}[t]
\centering
\begin{tikzpicture}[node distance=0.5cm, auto, every node/.style={font=\footnotesize}]
  \tikzstyle{block} = [rectangle, rounded corners, minimum width=3cm, minimum height=1cm, text centered, align=center, draw=black, fill=gray!10]
  
  \node[block] (1) {1. Data Collection\\
  \footnotesize (105 PDF reports + Open-Meteo \\\footnotesize Extracted using \texttt{pdfplumber})  };
  \node[block, right=0.8cm of 1] (2) {2. Preprocessing \& \\\footnotesize Feature Engineering\\\footnotesize (Lags, Weather, Structural)};
  \node[block, below=0.8cm of 2] (3) {3. EDA \& Pattern\\
  \footnotesize Analysis (Cyclone Ditwah)};
  \node[block, left=1cm of 3] (4) {4. Time-Series Diagnostics\\ \footnotesize(ADF, Granger, Event Window)};
  \node[block, below=0.8cm of 4] (5) {5. Modeling \& Evaluation\\ \footnotesize(Catalogue-Specific Model Comparison)};
  
  \draw[->, thick] (1) -- (2);
  \draw[->, thick] (2) -- (3);
  \draw[->, thick] (3) -- (4);
  \draw[->, thick] (4) -- (5);
  
\end{tikzpicture}
\caption{End-to-end methodology pipeline for Colombo Tea Auction price modeling.}
\label{fig:methodology_pipeline}
\end{figure}

\subsection{Data Collection}
Structured data was extracted from each weekly Forbes \& Walker PDF report 
using \texttt{pdfplumber}, a Python library specialized in table extraction 
from PDFs. A custom multi-table pipeline was developed to parse the reports 
and generate nine standardized CSV files per run. 
Table~\ref{tab:dataset} describes the main dataset overview.

In addition to market and price tables, weather information was collected 
from two sources. First, textual weather and crop condition descriptions 
were parsed directly from the \textbf{CROP AND WEATHER} section of each 
report, producing per-region crop intake direction 
(\texttt{text\_crop\_change}), matched weather condition terms 
(\texttt{text\_keywords}), and a composite regional severity score 
(\texttt{avg\_weather\_severity}). Second, daily weather data was retrieved 
from the Open-Meteo historical archive API for each tea-growing region and 
aggregated into 7-day windows; lagged features at 1, 2, and 3 weeks prior 
to each auction date were generated using \texttt{auction\_date} $-$ 
$7{\times}\text{lag}$ as the reference date, to capture delayed supply 
effects.

\begin{table}[tb]
\caption{Dataset Overview}
\centering
\setlength{\tabcolsep}{5pt}
\renewcommand{\arraystretch}{1.5}
\begin{tabular}{lccc}
\hline
\textbf{Property} & \textbf{Total Rows} & \textbf{Rows with Price} & \textbf{Missing Target} \\
\hline
Dust        & 2{,}205 & 2{,}171 & 34 (1.5\%) \\
High Grown  & 2{,}124 & 2{,}124 & 0 (0.0\%) \\
Low Grown   & 5{,}384 & 5{,}384 & 0 (0.0\%) \\
Off-Grade   & 2{,}520 & 2{,}118 & 402 (16.0\%) \\
\hline
\end{tabular}
\label{tab:dataset}
\end{table}

\subsection{Data Preprocessing}
A master analytical dataset is constructed by joining the price observations 
from the high-grown, low-grown, and off-grade/dust tables with sale-level 
context and region-aligned weather features using \texttt{sale\_id} as the 
primary key. The target variable \texttt{price\_mid\_lkr} is derived as 
$(\texttt{price\_lo\_lkr} + \texttt{price\_hi\_lkr}) / 2$, with a fallback 
to \texttt{price\_lo\_lkr} when only a lower bound is reported. The resulting 
modeling-ready dataset contains 12{,}233 rows.
Beyond the raw auction and weather variables, two categories of derived 
features were constructed.

\subsubsection{Temporal lags}
Lagged features were constructed for total precipitation (mm), mean temperature
($^\circ$C), and sunshine duration (seconds) at 1, 2, and 3 weeks before each
auction date. Each lag represents a 7-day window fetched from the Open-Meteo
API using \texttt{auction\_date} $- 7{\times}\text{lag}$ as the reference date.
This primary window targets short-term harvest, supply, and auction-market
disruption while preserving observations within the 105-week study period; it is
not intended to represent the full biological crop-to-auction pathway. To assess
longer weather histories, Granger tests were separately extended to maximum
cumulative lag orders of 4--8 weeks. Missing early-lag values were imputed using
the region-level median because precipitation distributions are right-skewed.

\subsubsection{Tea structural features} \texttt{elevation}, \texttt{grade}, and \texttt{tier} capture the physical and commercial 
hierarchy of each price observation, allowing catalogue-specific models to 
distinguish price behaviour across quality levels. \texttt{fx\_usd} provides 
the LKR/USD exchange rate consolidated from year-specific columns extracted 
from the PDF reports.

\subsection{Exploratory Data Analysis (EDA)}

Table~\ref{tab:descriptive_stats} summarizes the prices by its structural factors. Low Grown teas consistently command the highest average prices, followed by High Grown, Dust, and Off-Grade catalogues. This clear segmentation by elevation and catalogue type highlights the need for catalogue-specific modeling approaches in Sri Lanka tea auction price analysis.

Low Grown is the dominant catalogue in the dataset, both in size and variability, with 5,436 records, substantially higher than Dust, High Grown, and Off-Grade. Off-Grade further shows a notable data quality concern, with a 15.95\% rate of missing price values, which introduces potential bias in any aggregate comparisons unless handled in a catalogue-aware manner. In addition, categorical label completeness is uneven across catalogues. Structural categorical features such as \texttt{grade}, \texttt{tier}, and \texttt{catalogue} are not equally informative across catalogues, which explains why catalogue-specific modelling is often more stable.

\begin{table}[tb]
\centering
\caption{Descriptive Statistics of Tea Auction Prices by Market Catalogue}
\label{tab:descriptive_stats}
\setlength{\tabcolsep}{2pt}
\renewcommand{\arraystretch}{0.9}
\begin{tabular}{l l r r r r}
\toprule
\textbf{Group} & \textbf{Category} & \textbf{Mean(LKR)} & \textbf{Median(LKR)} & \textbf{SD(LKR)}\\
\midrule
\multirow{3}{*}{By Elevation} 
    & Low Grown  & 1,446.8 & 1,290.0 & 734.7 \\
    & High Grown & 1,057.6 & 1,100.0 & 268.4 \\
    & Medium Grown  & 859.9   & 900.0   & 189.8 \\
\midrule
\multirow{4}{*}{By Catalogue}
    & Low Grown     & 1,578.7 & 1,400.0 & 749.5  \\
    & High Grown    & 1,135.6 & 1,170.0 & 261.5 \\
    & Dust          & 984.4   & 985.0   & 207.3 \\
    & Off-Grade     & 792.2   & 730.0   & 169.3  \\
\midrule
\multicolumn{2}{l}{\textbf{Overall}} 
    & 1,255.8 & 1,120.0 & 623.7 \\
\bottomrule
\end{tabular}
{\footnotesize \textbf{Note:} The Low Grown ``By Elevation'' aggregates all tea grown below 600m, whereas the Low Grown ``By Catalogue'' is a specific commercial subdivision sorted by leaf size and style that excludes mixed categories like Off-Grade and Dust}
\end{table}

\subsection{Time-Series Diagnostics}

Auction prices are not \textit{a priori} stationary and applying regression based causality tests to non stationary series inflates type-I error. Therefore, we begin the analysis with three diagnostic steps.

\subsubsection{Stationarity}
The Augmented Dickey-Fuller (ADF) test confirmed that Low Grown price series is non-stationary in levels ($p{=}0.1087$) and required first-differencing. Conversely, High Grown, Off-Grade and Dust prices were found to be stationary at levels ($p<0.05$). Among weather variables, precipitation, temperature, and sunshine duration were generally stationary across all catalogues, ensuring valid Granger causality testing.

\subsubsection{Structural break.}
Cyclone Ditwah made landfall on 28 November 2025 and is a candidate structural break. An event window comparison of prices before and after landfall shows a short-lived upward shift in High Grown prices consistent with a supply shock (section~\ref{sec:ditwah_sec}).

\subsubsection{Granger causality}
\begin{figure}[tb]
\centering
\includegraphics[width=\columnwidth]{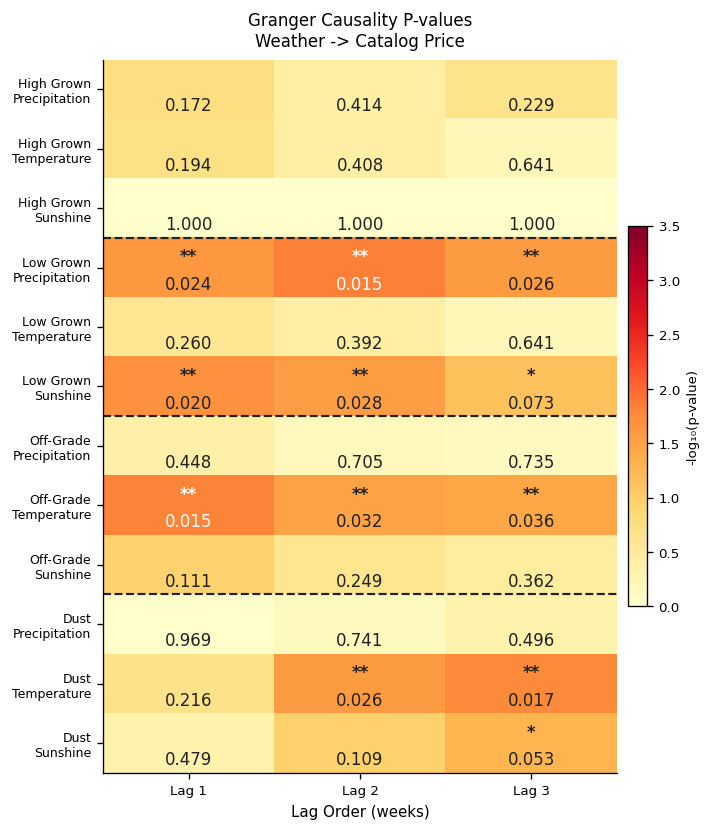}
\caption{Primary Granger test results for maximum cumulative lag orders
1--3. The heatmap is descriptive; inferential conclusions are based on
Benjamini--Hochberg-adjusted p-values.}
\label{fig:granger_heatmap}
\end{figure}

After ADF-guided differencing, the primary analysis tested each catalogue
$\times$ weather variable combination at maximum cumulative lag orders of 1--3
weeks (Figure~\ref{fig:granger_heatmap}). To assess longer weather histories,
the same procedure was separately extended to maximum cumulative lag orders
through 8 weeks. All weather--price conclusions are predictive associations in
the Granger sense, not structural causal claims.

Since Off-Grade and Dust teas are sourced from all three growing elevations,
no single weather region can fully represent their supply conditions. Rather
than selecting regions arbitrarily, we aligned these catalogues to the growing
areas most prominently featured in the broker reports, a choice that also sits
comfortably with established agronomic findings on how mid-to-high elevation
temperature conditions shape leaf quality in Sri Lanka~\cite{wijeratne1996vulnerability}. Alternative region mappings were not tested; however, this limitation has minimal impact given that all Granger findings are treated as exploratory.

\subsubsection{Multiple-comparisons Correction}
Benjamini-Hochberg correction~\cite{verma2014benjamini} was applied across
the 83 estimable tests in the extended 1--8 week lag family. Out of 96 planned
catalogue weather lag comparisons, 13 High Grown comparisons were not
estimable because the aggregated weather predictor series contained constant
values. Full test outputs are available in the
\gh{https://github.com/hesandism/data-analysis-for-tea-industry}{project repository};
results from the extended lag sensitivity analysis are reported in
Section~\ref{sec:granger}.

\subsection{Models and Training Protocols}

We evaluated four tree-based ensemble algorithms: \texttt{Random Forest}~\cite{breiman2001random},
\texttt{Gradient Boosting}~\cite{friedman2001greedy}, \texttt{XGBoost}~\cite{chen2016xgboost}, and
\texttt{LightGBM}~\cite{ke2017lightgbm}, under unified pooled and catalogue-specific
strategies. To assess value beyond short-term price autocorrelation, we also evaluated
Naive Persistence, where $\hat{p}_{t+1}=p_t$, and an AR(1) benchmark, where
$\hat{p}_{t+1}=\alpha+\beta p_t$, using only the current-week price within each
catalogue--grade--tier stream.

The target variable is \texttt{price\_next\_week}, constructed by shifting
\texttt{price\_mid\_lkr} forward by one auction week within each product stream.
All approaches were evaluated using $5$-fold chronological \texttt{TimeSeriesSplit}
cross-validation. For tree-based models, hyperparameters were selected through grid
search across eight candidate configurations per algorithm. This configuration count was chosen as a practical trade-off given the 105-week study window, and represents a limitation on tuning depth.

Median imputation was applied using \texttt{SimpleImputer} with the median
strategy. The imputer was fitted only on training data within each fold through
\texttt{scikit-learn}~\cite{pedregosa2011scikit} pipeline objects, preventing
test-set statistics from influencing imputation. Rows with missing forecasting
targets were excluded before training and evaluation; the effective Off-Grade
dataset therefore reflects records with valid price labels.

To quantify the incremental contribution of lagged weather, the selected best
tree-based model for each catalogue was re-evaluated after removing the nine
lagged precipitation, sunshine-duration, and temperature features. Full and
ablated models used identical chronological folds and all remaining features.
We report combined out-of-fold metrics and fold-wise RMSE mean and standard
deviation.

\section{Results and Analysis}

\subsection{Granger Causality Analysis}\label{sec:granger}

The primary analysis evaluated maximum cumulative lag orders of 1--3 weeks.
Among 33 estimable primary tests, 10 weather-price associations were nominally
significant at $p<0.05$. However, none survived Benjamini-Hochberg correction
across the full extended family of 83 estimable tests. These short-lag findings
are therefore interpreted as exploratory predictive signals rather than
confirmed causal effects.

To assess longer weather histories, the analysis was extended to maximum
cumulative lag orders of 4-8 weeks. Table~\ref{tab:long_lag_sensitivity}
summarizes the best raw result within each lag range. The only association that
survived correction was Dust temperature at maximum lag 4
($F=5.843$, $p=0.000310$, $q=0.0257$). This is a joint test of temperature
history across weeks 1-4, rather than an isolated fourth-week effect. No
maximum cumulative lag order from 5 through 8 weeks remained significant after
correction.

\begin{table*}[t]
\caption{Detailed Long-Lag Granger Sensitivity Results.}
\label{tab:long_lag_sensitivity}
\centering
\scriptsize
\setlength{\tabcolsep}{2pt}
\renewcommand{\arraystretch}{1.10}

\resizebox{0.7\linewidth}{!}{%
\begin{tabular}{llccc}
\hline
\textbf{Catalogue} & \textbf{Weather} &
\textbf{Best max.\ lag 1--3; raw $p$} &
\textbf{Best max.\ lag 4--8; raw $p$} &
\textbf{Min.\ FDR $q$} \\
\hline
High Grown & Precip. & 1; 0.1718 & NE & 0.3854 \\
High Grown & Temp.   & 1; 0.1936 & 5; 0.7464 & 0.4202 \\
High Grown & Sun.    & NE & NE & NE \\
Low Grown  & Precip. & 2; 0.0147 & 4; 0.0716 & 0.1898 \\
Low Grown  & Temp.   & 1; 0.2600 & 5; 0.5636 & 0.4877 \\
Low Grown  & Sun.    & 1; 0.0199 & 4; 0.0366 & 0.1898 \\
Off-Grade  & Precip. & 1; 0.4483 & 8; 0.5877 & 0.6645 \\
Off-Grade  & Temp.   & 1; 0.0152 & 8; 0.0484 & 0.1898 \\
Off-Grade  & Sun.    & 1; 0.1110 & 8; 0.1568 & 0.3291 \\
Dust       & Precip. & 3; 0.4957 & 8; 0.3479 & 0.6278 \\
Dust       & Temp.   & 3; 0.0167 & 4; 0.000310 & \textbf{0.0257} \\
Dust       & Sun.    & 3; 0.0527 & 8; 0.0499 & 0.2302 \\
\hline
\end{tabular}%
}

\vspace{2pt}
{\footnotesize \textit{Lag; $p$} reports the maximum cumulative lag order with
the smallest raw p-value within the stated group. Min.\ $q$ is the minimum
Benjamini--Hochberg-adjusted p-value across all estimable lag orders 1--8.
NE = not estimable because the aggregated predictor series contained constant
values. A maximum lag of $k$ jointly tests weather information from weeks 1
through $k$.}
\end{table*}

\subsection{Model Performance}

\subsubsection{Evaluation Metrics}

Model performance is assessed using four metrics computed from combined out-of-fold
predictions across five chronological \texttt{TimeSeriesSplit} folds.

\noindent \textbf{Root Mean Squared Error (RMSE):} The primary ranking criterion,
measuring prediction error in the original LKR scale. It penalizes large errors and is
sensitive to high price variance.

\noindent \textbf{Mean Absolute Error (MAE):} The average absolute deviation per lot
in LKR, providing an interpretable measure of typical error.

\noindent \textbf{Coefficient of Determination ($R^2$):} Measures the proportion of
price variance captured relative to a mean-prediction baseline. A positive $R^2$
indicates predictive value, while values $\leq 0$ do not outperform the catalogue mean.

\noindent \textbf{Mean Absolute Percentage Error (MAPE):} The mean absolute error
expressed as a percentage of the actual price. Since a fixed LKR error produces a
larger percentage error for lower-priced catalogues, MAPE should be interpreted
alongside RMSE rather than compared directly across catalogues.

\subsubsection{Catalogue-Specific Models}
Table~\ref{tab:baselines} compares price-only baselines with the four tree-based
models. Among the tree-based models, \texttt{LightGBM} achieved the lowest RMSE for
High Grown and Off-Grade, while \texttt{XGBoost} performed best for Dust. However,
Naive Persistence was the best overall method for Low Grown, achieving an RMSE of
$79.76$ LKR and an $R^2$ of $0.9896$, compared with \texttt{LightGBM}'s RMSE of
$145.43$ LKR and $R^2$ of $0.9653$.

For High Grown, Off-Grade, and Dust, the best tree-based model improved on Naive
Persistence by $33.31\%$, $45.78\%$, and $52.42\%$ in RMSE, respectively. Thus,
tree-based models add forecasting value for these catalogues, whereas Low Grown's
high one-week predictability is primarily explained by short-term price persistence.

\begin{table}[t]
\caption{One-Week-Ahead Forecast Performance: Baselines and Tree-Based Models.}
\label{tab:baselines}
\footnotesize
\setlength{\tabcolsep}{3pt}
\centering
\renewcommand{\arraystretch}{1.3}
\begin{tabular}{llrrrr}
\toprule
\textbf{Catalogue} & \textbf{Method} & \textbf{RMSE} & \textbf{MAE} &
\textbf{MAPE (\%)} & \textbf{$R^2$} \\
\midrule
\multirow{6}{*}{High Grown}
 & Naive Persistence & 250.44 & 169.93 & 14.48 & $-$0.241 \\
 & AR(1) & 209.11 & 148.83 & 12.79 & 0.135 \\
 & XGBoost & 170.14 & 100.51 & 8.37 & 0.427 \\
 & \textbf{LightGBM} & \textbf{167.02} & 101.64 & 8.49 & \textbf{0.448} \\
 & Gradient Boosting & 171.51 & 102.60 & 8.52 & 0.418 \\
 & Random Forest & 177.23 & 103.12 & 8.57 & 0.378 \\
\midrule
\multirow{6}{*}{Low Grown}
 & \textbf{Naive Persistence} & \textbf{79.76} & 34.80 & 1.91 & \textbf{0.990} \\
 & AR(1) & 80.91 & 38.89 & 2.21 & 0.989 \\
 & XGBoost & 152.31 & 77.52 & 4.15 & 0.962 \\
 & LightGBM & 145.43 & 74.87 & 4.11 & 0.965 \\
 & Gradient Boosting & 149.22 & 76.50 & 4.20 & 0.964 \\
 & Random Forest & 151.97 & 77.30 & 4.21 & 0.962 \\
\midrule
\multirow{6}{*}{Off-Grade}
 & Naive Persistence & 188.39 & 140.64 & 16.76 & $-$0.664 \\
 & AR(1) & 145.06 & 121.44 & 14.90 & 0.014 \\
 & XGBoost & 103.83 & 80.60 & 10.00 & 0.495 \\
 & \textbf{LightGBM} & \textbf{102.15} & 77.37 & 9.54 & \textbf{0.511} \\
 & Gradient Boosting & 107.33 & 82.17 & 10.18 & 0.460 \\
 & Random Forest & 104.74 & 81.22 & 10.12 & 0.486 \\
\midrule
\multirow{6}{*}{Dust}
 & Naive Persistence & 241.29 & 187.78 & 18.90 & $-$0.621 \\
 & AR(1) & 189.31 & 145.17 & 14.76 & 0.002 \\
 & \textbf{XGBoost} & \textbf{114.82} & 76.36 & 7.46 & \textbf{0.633} \\
 & LightGBM & 117.48 & 78.12 & 7.64 & 0.616 \\
 & Gradient Boosting & 119.87 & 85.08 & 8.38 & 0.600 \\
 & Random Forest & 116.18 & 75.37 & 7.36 & 0.624 \\
\midrule
\multirow{6}{*}{Pooled (Unified)}
 & Naive Persistence & 176.98 & 106.44 & 9.98 & 0.922 \\
 & AR(1) & 175.33 & 108.68 & 10.17 & 0.923 \\
 & XGBoost & 142.18 & 83.05 & 6.74 & 0.950 \\
 & \textbf{LightGBM} & \textbf{139.17} & 83.52 & 6.88 & \textbf{0.952} \\
 & Gradient Boosting & 141.27 & 89.69 & 7.51 & 0.950 \\
 & Random Forest & 144.97 & 83.09 & 6.74 & 0.948 \\
\bottomrule
\end{tabular}
\vspace{2pt}
{\footnotesize \textbf{Bold} = lowest RMSE within each catalogue.
Naive Persistence: $\hat{y}_{t+1}=y_t$. AR(1): first-order price
autoregression. All metrics are computed from combined OOF predictions
across $5$ chronological folds.}
\end{table}

\subsubsection{Weather-Feature Ablation and Fold Variability}
Table~\ref{tab:weather_ablation} isolates the contribution of lagged weather
features. Removing lagged weather increased out-of-fold RMSE by $1.10$ LKR for
High Grown, $5.03$ LKR for Low Grown, $2.19$ LKR for Off-Grade, and $4.78$ LKR
for Dust. These gains indicate modest incremental value from lagged weather,
with the clearest improvements observed for Low Grown and Dust. However, the
gains are small relative to fold-level RMSE variation and should be interpreted
cautiously. Low Grown's full machine-learning model also remains less accurate
than the Naive Persistence baseline.

\begin{table}[t]
\caption{Lagged Weather Feature Ablation and Fold-Level RMSE Variability.}
\label{tab:weather_ablation}
\footnotesize
\setlength{\tabcolsep}{8pt}
\centering
\renewcommand{\arraystretch}{1.3}
\begin{tabular}{lrrrr}
\toprule
\textbf{Catalogue} & \textbf{Full} & \textbf{No Lag} &
\textbf{$\Delta$} & \textbf{Fold RMSE} \\
\midrule
High Grown & 167.02 & 168.12 & +1.10 & $160.97 \pm 49.78$ \\
Low Grown & 145.43 & 150.47 & +5.03 & $142.48 \pm 32.58$ \\
Off-Grade & 102.15 & 104.33 & +2.19 & $100.87 \pm 17.97$ \\
Dust & 114.82 & 119.60 & +4.78 & $108.13 \pm 43.16$ \\
\bottomrule
\end{tabular}
\vspace{2pt}
{\footnotesize \textbf{Note: }Full and No Lag are combined OOF RMSE values. Positive
$\Delta$ indicates lower error when lagged weather features are retained.
Fold RMSE is reported as mean $\pm$ standard deviation across $5$ chronological folds.}
\end{table}

\subsubsection{Unified Pooled Model}
In the pooled setting, \texttt{LightGBM} achieved the lowest RMSE of $139.17$ LKR and
an $R^2$ of $0.9516$. It improved on Naive Persistence by $21.36\%$ in RMSE, indicating
that the richer pooled feature set adds predictive value beyond short-term price
persistence. However, the pooled result is influenced by the larger Low Grown
catalogue and should be interpreted alongside the catalogue-specific results.

\section{Key Findings}

\subsection{Short-Term Market Response to Extreme Weather: Evidence from Cyclone Ditwah}\label{sec:ditwah_sec}
\begin{figure*}[!htb]
\centering
\includegraphics[width=\linewidth]{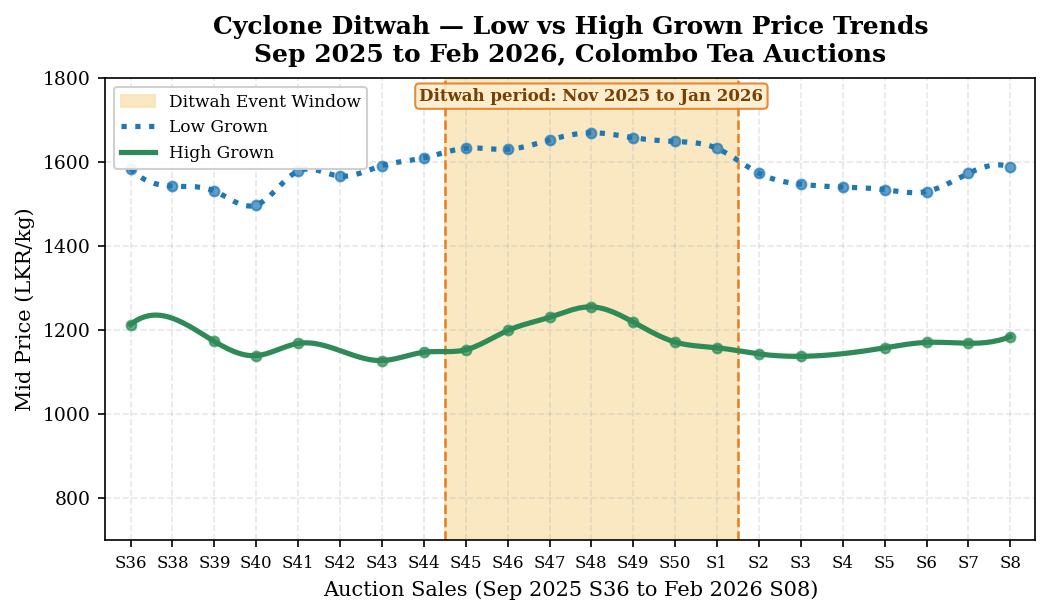}
\caption{Event-window analysis of mean tea auction prices around Cyclone Ditwah (November 2025-January 2026). The shaded region denotes the primary cyclone impact window. High-grown teas display the strongest short-term price spike.}
\label{fig:ditwah_event_window}
\end{figure*}

Cyclone Ditwah struck Sri Lanka's eastern coast on 28 November 2025, bringing heavy flooding and landslides that severely affected many tea growing areas\footnote{\url{https://www.aljazeera.com/news/2025/12/10/like-wastelands-sri-lanka-tea-plantations-suffer-cyclone-ditwahs-wrath}}. Looking at the event window around the cyclone (Figure~\ref{fig:ditwah_event_window}), both Low Grown and High Grown mid-prices move upward into a mid-window peak and then soften after the window closes, suggesting a temporary event-period price pressure rather than a permanent level shift.

Notably, the Cyclone Ditwah impact on High Grown prices occurs quickly and fades rapidly. Even though this event-window analysis reveals a visible short-term price 
adjustment consistent with a supply shock, a single cyclone cannot establish a generalizable pattern. Nonetheless, the directional finding aligns with agronomic expectations and motivates closer monitoring of extreme weather events in future auction series as more events accumulate

\subsection{Clean and Reproducible Tea Auction Dataset}

To support reproducible forecasting, we construct a cleaned 
tea auction dataset from raw PDF reports and release it via public repositories on \hf{https://huggingface.co/datasets/hesandism/colombo-tea-auction-prices}{huggingface} and \gh{https://github.com/anonymous-research-tea/tea-auction-dataset.git}{github}.
The dataset integrates sale-level, price-level, and weather 
variables using \texttt{sale\_id} as the primary key, with 
rule-based mapping between tea categories and weather regions.

The final dataset contains 12{,}233 records with 26 features across 105 sales from November 2023 to March 2026. This dataset serves as a reusable research artifact, enabling reproducibility and standardized analysis of Sri Lankan tea auction data. Our \gh{https://github.com/hesandism/data-analysis-for-tea-industry}{code} is also publicly available.

\section{Discussion}
\subsection{Weather Impact on Tea Prices}
Weather signals were limited and catalogue-specific. Although several short-lag
weather--price associations were nominally significant, none survived
Benjamini--Hochberg correction across the extended lag family. The only robust
result emerged in the 1--8 week sensitivity analysis: Dust prices were
Granger-predicted by cumulative temperature history through maximum lag 4
($F=5.843$, $p=0.000310$, $q=0.0257$; Section~\ref{sec:granger}). This is a
joint test of temperature information across weeks 1--4, rather than an
isolated fourth-week effect. No maximum cumulative lag orders from 5 through
8 weeks remained significant after correction.

The event-window around Cyclone Ditwah suggests a temporary High Grown price
adjustment. However, this single-event pattern does not establish a
generalizable weather effect and should be interpreted as descriptive evidence
for continued monitoring of extreme-weather shocks.

The lagged-weather ablation provides modest evidence of incremental forecasting
value beyond the remaining model features. Removing lagged weather increased
out-of-fold RMSE by $5.03$ LKR for Low Grown, $2.19$ LKR for Off-Grade, and
$4.78$ LKR for Dust, while the High Grown change was only $1.10$ LKR. These
differences are small relative to fold-level variation and should therefore be
treated as exploratory predictive gains rather than decisive weather effects.

For Low Grown, lagged weather improved the machine-learning model modestly, but
Naive Persistence remained substantially more accurate. Thus, short-term price
autocorrelation dominates one-week-ahead Low Grown forecasting despite the small
incremental contribution of lagged weather variables.

\subsection{Rationale for Catalogue-Specific Modeling}
The tea market comprises High Grown, Low Grown, Off-Grade, and Dust catalogues with
distinct price levels and short-term dynamics. Although the pooled model performs
strongly overall, its accuracy is influenced by the larger Low Grown catalogue and can
mask catalogue-specific behaviour.

No single method was optimal across all catalogues. \texttt{LightGBM} achieved the
lowest RMSE for High Grown and Off-Grade, while \texttt{XGBoost} performed best for
Dust. In contrast, Naive Persistence was the best method for Low Grown, indicating
that its one-week-ahead predictability is dominated by short-term price continuity.
Catalogue-specific evaluation is therefore necessary to identify where richer
machine-learning models add value and where a simple operational baseline is more
appropriate.

The pooled \texttt{LightGBM} model remains useful for market-level forecasting;
however, it should be interpreted alongside catalogue-specific results rather than as
a replacement for them.

\subsection{Limits of One-Week-Ahead Forecasting}

High Grown and Dust prices exhibit substantial week-ahead variation driven by
auction-day factors such as buyer participation, bidding behaviour, and blend-maker
decisions that are not captured in pre-auction data. Low Grown prices show stronger
week-to-week continuity; however, Naive Persistence outperformed the machine-learning
models, indicating that its high predictability largely reflects price autocorrelation.

Future work should incorporate auction-day signals, such as buyer activity or
real-time bidding data, to improve forecasts for High Grown and Dust catalogues.

\section{Conclusion}
This study establishes that Colombo Tea Auction prices are not a monolithic series but rather a collection of distinct market catalogues with different price dynamics and forecasting behaviour. Extended Granger analysis across cumulative lag orders of 1--8 weeks identified one association that remained significant after Benjamini--Hochberg correction: temperature history across weeks 1--4 Granger-predicted Dust prices ($F=5.843$, $p=0.000310$, $q=0.0257$). Other nominal weather--price associations across Low Grown, Off-Grade, and Dust catalogues did not survive correction and should be interpreted as exploratory predictive signals rather than confirmed weather effects. The Cyclone Ditwah event-window analysis further suggests a short-lived High Grown price response, although this single-event observation is not sufficient to establish a generalizable extreme-weather effect.

From a forecasting perspective, performance differed across catalogues.
\texttt{LightGBM} achieved the lowest RMSE for High Grown and Off-Grade, while
\texttt{XGBoost} performed best for Dust. However, Low Grown prices were forecast most
accurately by Naive Persistence, with an RMSE of $79.76$ LKR and an $R^2$ of $0.9896$.
This indicates that strong short-term price autocorrelation, rather than the current
richer feature set, dominates one-week-ahead Low Grown forecasting.

Lagged-weather ablation showed modest incremental forecasting gains for Low
Grown, Off-Grade, and Dust, but little benefit for High Grown; these gains should
be interpreted cautiously given fold-level variability.

In the pooled setting, \texttt{LightGBM} ranked first with an RMSE of $139.17$ LKR and
an $R^2$ of $0.9516$, improving on Naive Persistence by $21.36\%$ in RMSE. These
results support catalogue-specific evaluation and the practical use of simple
baselines alongside machine-learning models.

{\footnotesize
\bibliographystyle{IEEEtranN}
\bibliography{references}
}

\end{document}